\documentclass[aps,preprint,amsmath,amssymb,nofootinbib]{revtex4}

\usepackage{hyperref}
\usepackage{graphicx}
\usepackage{slashed}

\begin{document}

\title{Earth Matter Effect on Democratic Neutrinos}


\author{Dmitry Zhuridov}%
\email{dmitry.zhuridov@gmail.com}%
\affiliation{Institute of Physics, University of Silesia,
Katowice, PL 40-007, Poland;}
\affiliation{Department of Physics and Astronomy, Wayne State University,
Detroit, MI 48201, USA}

\date{\today} 

\begin{abstract}

The neutrino propagation through the Earth is investigated in the framework of the democratic neutrino theory. In this theory the neutrino mixing angle $\theta_{13}$ is approximately determined, which allows one to make a well defined neutrino oscillogram driven by the 1-3 mixing in the matter of the Earth. Significant differences in this oscillogram from the case of models with relatively small $\theta_{13}$ are discussed. 

\end{abstract}

\maketitle

\section{Introduction}

The Democratic Neutrino Theory (DNT), which was introduced in Refs.~\cite{Zhuridov:2013eqa,Zhuridov:2013ika,Zhuridov:2014vfa}, 
is based on a simple $S_3$ symmetric leading order democratic mass matrix, which allows to make certain predictions 
on the neutrino masses and mixing. In particular, a generic neutrino mass spectrum in DNT with small perturbations 
can be approximately written as (small deviation of $m_3$ from $2m$ is irrelevant)
\begin{eqnarray}
	\{ m, m+\delta m, 2m \},
\end{eqnarray}
where $m\approx0.03$~eV is the absolute neutrino mass scale, which is determined from the atmospheric neutrino oscillation data, 
and $\delta m \ll m$ is a perturbation in the democratic neutrino mass spectrum. In this model the mixing angles 
can be approximately determined as $\theta_{12}\approx30^\circ$, $\theta_{23}\approx45^\circ$ and $\theta_{13}\approx35.3^\circ$ 
(in the standard parametrization). These predictions agree with the present solar, atmospheric and accelerator neutrino data. 
However the alternative explanations with respect to the Conventional Neutrino Theory (CNT)~\cite{PDG2012} of many neutrino results 
are required~\cite{Zhuridov:2014vfa}. In particular, the suppression of the atmospheric $\nu_e\to\nu_{\mu,\tau}$ oscillations, 
which was observed by the Super-Kamiokande~\cite{Ashie:2004mr,Ashie:2005ik}, is explained in CNT by small $|U_{e3}|^2$. 
It was proposed in Ref.~\cite{Zhuridov:2014vfa}  that in DNT this suppression may take place due to the Earth's matter effect. 
In this paper we investigate this possibility in more detail. 

Another essential difference between DNT and CNT is in explanation of the large amplitude of the atmospheric $\nu_\mu$ oscillations, 
which was observed by the Super-Kamiokande and MINOS~\cite{Adamson:2011ig,Adamson:2013whj}. In CNT this close to unity amplitude is explained by the large $\nu_\mu\to\nu_\tau$ oscillations 
due to the large values of $|U_{\mu3}|$ and $|U_{\tau3}|$, where each of them is about $1/\sqrt{2}$. However in DNT this result is explained 
by the sum of $\nu_\mu\to\nu_\tau$ and $\nu_\mu\to\nu_e$ oscillations, where they both are significant.

Finally,  
in DNT the fluxes of incoherent massive neutrino eigenstates at the Earth correspond to the mass composition of 
the electron neutrino at production in the core of the Sun. However CNT assumes either that these fluxes correspond 
to the mass composition of neutrino at the last scattering surface or neutrino coherence at the Earth. 
First possibility does not work since the interactions of neutrinos with the matter of the Sun do not change the lepton number. 
Second possibility is not realistic due to the separation of the neutrino wave packets. Hence the explanation 
of solar neutrinos in CNT is in question.

We should stress that due to these differences in explanation of the particular neutrino phenomena in DNT and CNT 
the results of conventional global analysis of the neutrino data can not be applied to DNT directly. This paper is one of the steps in the direction to an adequate global analysis, which is required for accurate verification of DNT.

We notice a strong hierarchy $\Delta m_{21}^2 \ll \Delta m_{31}^2$ between the two neutrino mass splittings $\Delta m_{31}^2\equiv\Delta m^2=3m^3$ and $\Delta m_{21}^2\equiv\Delta\mu^2\approx2m\delta m$ in DNT with  tiny perturbations, which significantly simplifies the investigation of the neutrino propagation through the Earth. Moreover the given value of the 1-3 mixing angle removes the major uncertainty in the neutrino oscillogram, which is investigated in the next section.

\section{Neutrino oscillograms of the Earth:  DNT vs. CNT}

The evolution of the neutrino state over a finite distance from $x_0$ to $x$ can be described using the evolution matrix $S(x,x_0)$~\cite{Akhmedov:2006hb,Akhmedov:2008qt,Akhmedov:2012ah}. The matrix $S(x) = S(x,0)$ satisfies the same evolution equation as the state vector $\nu_f = (\nu_e,\nu_\mu,\nu_\tau)^T$, namely,
\begin{eqnarray}
	i \frac{dS(x)}{x}	=	\hat H(x) S(x).
\end{eqnarray}
In this equation the Hamiltonian can be written as\footnote{In sections~\ref{Section:A} and \ref{Section:B} a symmetric Hamiltonian (with $H_{11}=-H_{22}=-\cos2\theta_{13}\Delta m^2/4E+V/2$) for $2\nu$ system is used, which differs from Eq.~\eqref{Eq:Ham} by a term proportional to the unit matrix, see Ref.~\cite{Akhmedov:2006hb}.}

\begin{eqnarray}\label{Eq:Ham}
	\hat H(x) 	= 	\frac{U M^2 U^\dag}{2 E_\nu}	+	\hat V(x),
\end{eqnarray}
where $E_\nu$ is the neutrino energy, $M^2= \rm{diag}(0,\Delta m_{21}^2,\Delta m_{31}^2)$ is the neutrino splitting matrix; 
\begin{eqnarray}
	U	=	R_{23} I_\delta R_{13} I_{-\delta} R_{12}
\end{eqnarray}
is the neutrino mixing matrix, where $R_{ij} = R_{ij} (\theta_{ij})$ is the Euler rotation matrix, and $I_\delta = \text{diag}(1, 1, e^{i\delta})$ with $CP$-violating Dirac phase $\delta$; and $\hat V(x)= \rm{diag}(V(x),0,0)$ is the matrix of matter-induced neutrino potentials with
\begin{eqnarray}
	V(x)=\sqrt{2}G_F N_e(x),
\end{eqnarray}
where $G_F$ is the Fermi constant, and $N_e(x)$ is the electron number density.

In the propagation basis $\tilde\nu= (\nu_e,\tilde\nu_2,\tilde\nu_3)^T$, which can be defined through the transformation $\nu_f = \tilde U  \tilde\nu$ with $\tilde U = R_{23} I_\delta$, the evolution matrix can be written as
\begin{eqnarray}
	\tilde S(x) 	=	\tilde U^\dag  S(x) \tilde U,
\end{eqnarray}
and parametrized as
\begin{eqnarray}
	\tilde S \equiv \tilde S(L) 	= 
\left(
\begin{array}{ccc}
 A_{ee} & A_{e\bar2}  & A_{e\bar3}  \\
 A_{\bar2e} & A_{\bar2\bar2}  & A_{\bar2\bar3}  \\
 A_{\bar3e} & A_{\bar3\bar2}  & A_{\bar3\bar3}  
\end{array}
\right),
\end{eqnarray}
where $L$ is the total length of the neutrino trajectory.

In the DNT with small perturbations the 1-2 mass splitting can be neglected to a good precision (not only for high energies $E_\nu \gtrsim 1$~GeV as in CNT). In this approximation $A_{e\bar2} = A_{\bar2e} = A_{\bar2\bar3} = A_{\bar3\bar2} = 0$, $A_{\bar2\bar2} = 1$ and the evolution matrix in the flavor basis can be rewritten as~\cite{Akhmedov:2006hb}
\begin{eqnarray}
	\tilde S 	= 
\left(
\begin{array}{ccc}
 A_{ee} & s_{23}A_{e\bar3}  & c_{23}A_{e\bar3}  \\
 s_{23}A_{\bar3e} & c_{23}^2A_{\bar2\bar2} + s_{23}^2A_{\bar3\bar3}  & -s_{23}c_{23}(A_{\bar2\bar2} -A_{\bar3\bar3}) \\
 c_{23}A_{\bar3e} & -s_{23}c_{23}(A_{\bar2\bar2} -A_{\bar3\bar3})  & s_{23}^2A_{\bar2\bar2} + c_{23}^2A_{\bar3\bar3} 
\end{array}
\right),
\end{eqnarray}
where $c_{ij}=\cos\theta_{ij}$ and $s_{ij}=\sin\theta_{ij}$. 
Using the value $\theta_{23}\approx\pi/4$ of DNT with tiny perturbations, the neutrino oscillation probabilities~\cite{Akhmedov:2006hb,Akhmedov:2012ah}, which are relevant to the atmospheric neutrino fluxes, can be written as
\begin{eqnarray}
	&& P(\nu_e\to\nu_e) = 1 - P_A,	\\
	&& P(\nu_e\to\nu_\mu) = P(\nu_\mu\to\nu_e) \approx \frac{1}{2} P_A, \\
	&& P(\nu_\mu\to\nu_\mu) \approx \frac{1}{2} - \frac{1}{4} P_A + \frac{1}{2} \sqrt{1-P_A} \cos{\phi_X},  \\
	&& P(\nu_e\to\nu_\tau) = P(\nu_\tau\to\nu_e) \approx \frac{1}{2} P_A, \\
	&& P(\nu_\mu\to\nu_\tau) = P(\nu_\tau\to\nu_\mu) \approx \frac{1}{2} - \frac{1}{4} P_A - \frac{1}{2} \sqrt{1-P_A} \cos{\phi_X},
\end{eqnarray}
where 
\begin{eqnarray}\label{eq:PA_def}
	 P_A = |A_{e\bar3}|^2
\end{eqnarray}
is the two-neutrino transition $\nu_e\leftrightarrow\nu_{\mu,\tau}$ probability, and $\phi_X  = \arg \left[ A_{\bar2\bar2} A_{\bar3\bar3}^* \right]$.

In our calculations we use the electron density profile $N(r)=N_e(r)/N_A$ (where $r$ is the distance from the center of the Earth, and $N_A$ is the Avogadro constant), which was derived in Ref.~\cite{Lisi:1997yc} using the PREM model~\cite{Dziewonski:1981xy} for the matter density distribution in the Earth. The correspondent neutrino potential is symmetric with respect to the midpoint of the trajectory
\begin{eqnarray}
	 V(x)=V(L-x),
\end{eqnarray}
where $L=2R\cos\Theta_\nu$ is the length of the neutrino trajectory, which corresponds to a nadir angle $\Theta_\nu$. 
For the mantle-only crossing trajectories ( $33.1^\circ<\Theta_\nu<90^\circ$) $V(x)$ can be considered as one matter layer with relatively weakly varying density, while for the core-crossing trajectories  ( $0^\circ<\Theta_\nu<33.1^\circ$) $V(x)$ can be considered as three matter layers of this type. For each of these layers the neutrino potential $V(x)$ along a given trajectory can be written as
\begin{eqnarray}
	V(x)=	\bar V	+	\Delta V(x),
\end{eqnarray}
where $\bar V$ is a constant term, and $\Delta V(x)$ is a small perturbation.

\subsection{Mantle-only crossing trajectories}\label{Section:A}

Consider the neutrino trajectories, which cross only the Earth mantle ( $33.1^\circ<\Theta_\nu<90^\circ$). Using a perturbation theory in $\Delta V$ the two-neutrino transition probability in Eq.~\eqref{eq:PA_def} can be rewritten as~\cite{Akhmedov:2006hb}
\begin{eqnarray}\label{eq:PA}
	P_A =	\left(	\cos\varepsilon \sin2\bar\theta \sin\phi	+	\sin\varepsilon \cos2\bar\theta	\right)^2,
\end{eqnarray}
where $\bar\theta=\theta_{13}^m(\bar V)$ is the mixing angle in matter, $\phi	=	\phi_{13}^{m\bar V}(L)$, and $\varepsilon = \sin2\bar\theta \Delta I$ with
\begin{eqnarray}
	\Delta I = 	\frac{1}{2} \int_{-L/2}^{L/2}		\Delta V\left(z+\frac{L}{2}\right)	\cos[\phi_{13}^{m\bar V}(z)]dz.
\end{eqnarray}
We have used the expressions~\cite{Akhmedov:2012ah}
\begin{eqnarray}
	\cos2\theta_{13}^m		=	\frac{\cos2\theta_{13}\Delta-V}{	\sqrt{(\cos2\theta_{13}\Delta-V)^2	+	\Delta^2\sin^22\theta_{13}}},
\end{eqnarray}
\begin{eqnarray}
	\phi_{13}^{m\bar V}(x)	=	x	\sqrt{(\cos2\theta_{13}\Delta-\bar V)^2	+	\Delta^2\sin^22\theta_{13}},
\end{eqnarray} 
where $\Delta\equiv\Delta m_{31}^2/(2E_\nu)$, and averaged the neutrino potential along the trajectory as
\begin{eqnarray}
	\bar V(\Theta_\nu)	=	\frac{1}{L(\Theta_\nu)}	\int_0^{L(\Theta_\nu)}	V \left[r(x,\Theta_\nu)\right]dx,
\end{eqnarray} 
where
\begin{eqnarray}
	r(x,\Theta_\nu) = 	\sqrt{		\left( \frac{L(\Theta_\nu)}{2} \tan\Theta_\nu \right)^2	+	\left(	\frac{L(\Theta_\nu)}{2} - x \right)^2	}
\end{eqnarray} 
is the distance from the Earth's center to the point $x$ of the neutrino trajectory.

\subsection{Core-crossing trajectories}\label{Section:B}

In case of the core-crossing neutrino trajectories ( $0^\circ<\Theta_\nu<33.1^\circ$) the two-neutrino transition probability can be found as~\cite{Akhmedov:2006hb}
\begin{eqnarray}
	P_A =	|S_{12}|^2,
\end{eqnarray}
using the factorized evolution matrix $S=S_1^T S_2 S_1$, where the matrix $S_1$ ($S_2$) corresponds to the neutrino evolution in the appropriate region of the Earth's mantle (core). In particular, 
\begin{eqnarray}
	S_2	=		\cos\varepsilon_2  \bar S_2	-	i  \sin\varepsilon_2  	
				 \left( \begin{array}{cc}
    \sin2\bar\theta_2	 &	  \cos2\bar\theta_2	 \\ 
    \cos2\bar\theta_2	 &	- \sin2\bar\theta_2  \\ 
  \end{array}
				 \right),
\end{eqnarray}
where $\varepsilon_2 = \sin2\bar\theta_2 \Delta I_2$  and
\begin{eqnarray}
	\bar S_2	= 	
				 \left( \begin{array}{cc}
    \cos\phi_2+i \cos2\bar\theta_2\sin\phi_2	 &	 -i\sin2\bar\theta_2\sin\phi_2	 \\ 
   -i\sin2\bar\theta_2\sin\phi_2	 &	 \cos\phi_2-i \cos2\bar\theta_2\sin\phi_2  \\ 
  \end{array}
				 \right)
\end{eqnarray}
with $\phi_2 = \phi_{13}^{m\bar V_2}(L - 2L_1)$ and $\Delta I_2 = \Delta I (\bar V \to \bar V_2,L \to L - 2L_1)$.

Following Ref.~\cite{Akhmedov:2006hb} we approximate the density profile within each mantle layer by a linear function
\begin{eqnarray}
	V_1(z) = \bar V_1 + \Delta V_1(z), 	\qquad	\Delta V_1(z) \approx 	V_1^\prime \frac{z}{L_1},
\end{eqnarray}
where $L_1$ is the length of the trajectory within one mantle layer, $\bar V_1=\bar V(L \to L_1)$ and $z=x-L_1/2$.  At first-order in $\varepsilon_1 = \sin2\bar\theta_1 \Delta J_1$:
\begin{eqnarray}
	S_1	\approx	
				 \left( \begin{array}{cc}
    \cos\phi_1+i \cos2\bar\theta_1\sin\phi_1	 &	 \sin2\bar\theta_1(-i\sin\phi_1-\Delta J_1)	 \\ 
    \sin2\bar\theta_1(-i\sin\phi_1+\Delta J_1)	 &	 \cos\phi_1-i \cos2\bar\theta_1\sin\phi_1  \\ 
  \end{array}
				 \right),
\end{eqnarray}
where $\bar\theta_1=\theta_m(\bar V_1)$, $\phi_1 = \phi_{13}^{m\bar V_1}(L_1)$  and
\begin{eqnarray}
	\Delta J_1 = 	V_1^\prime L_1 	\frac{\sin\phi_1-\phi_1\cos\phi_1}{4\phi_1^2}.
\end{eqnarray}

\subsection{Neutrino oscillograms}

\begin{center}
 \begin{figure}[tb]
 \centering
 	\includegraphics[width=0.42\textwidth]{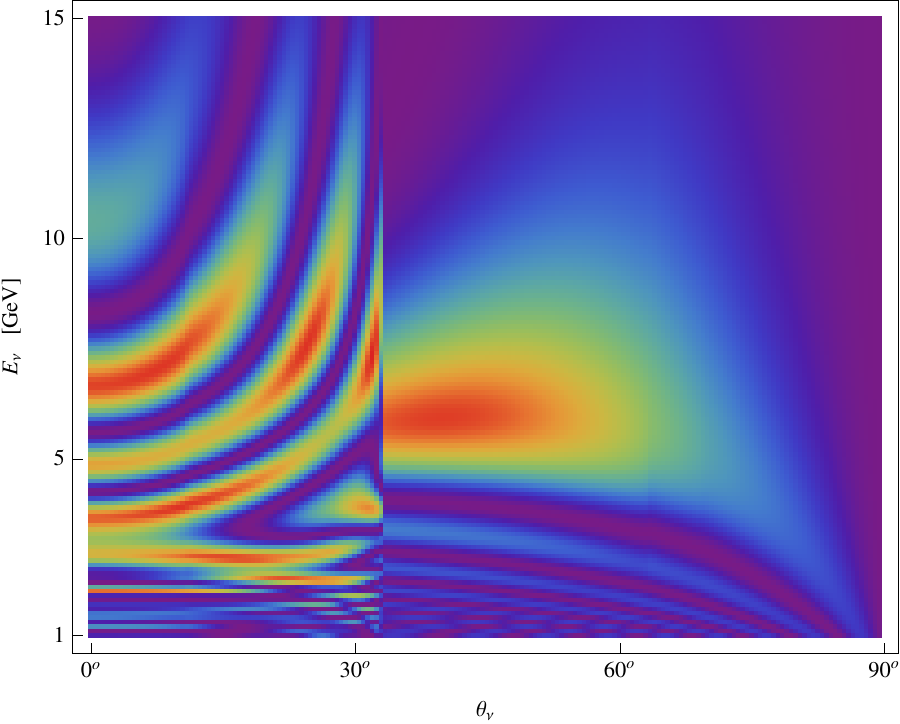}
	\includegraphics[width=0.5\textwidth]{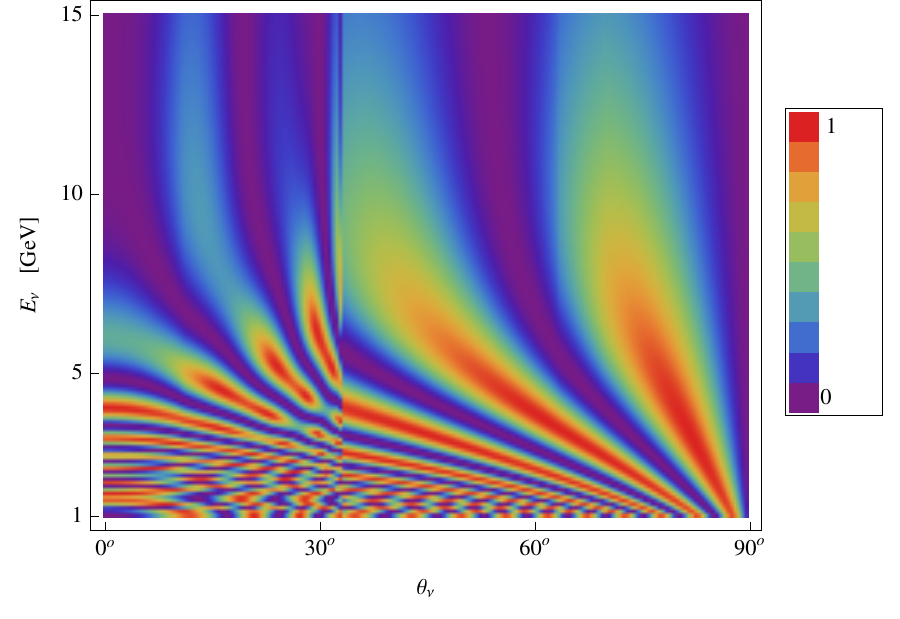}
   \caption{Neutrino oscillograms  calculated within:  CNT with $\sin^22\theta_{13}=0.1$ ({\it left}) and DNT ({\it right}). Shown are the values of $P_A$ in the plane of the nadir angle $\Theta_\nu$ and the neutrino energy $E_\nu$.}
   \label{Fig:oscillograms}
 \end{figure}
\end{center}

The neutrino oscillograms 
calculated in the considered approximation of a matter layer with weakly varying density are shown in Fig.~\ref{Fig:oscillograms} 
for the range of neutrino energies from 1~GeV to 15~GeV. (For lower neutrino energies the matter effect on atmospheric neutrinos 
is mainly determined by the electron number density within one oscillation length under the detector~\cite{Lobanov}.) 
The oscillogram in the left is calculated within CNT with $\sin^22\theta_{13}=0.1$. Its part for the mantle-only crossing trajectories 
($\Theta_\nu\gtrsim33^\circ$) accurately reproduces the corresponding result of Ref.~\cite{Akhmedov:2006hb}. Some differences with the result of Ref.~\cite{Akhmedov:2006hb} for the core-crossing trajectories require additional consideration. The oscillogram in the right is calculated within DNT, in which the value of $\theta_{13}$ is significantly larger ($\sin^22\theta_{13}=8/9$). 
\vspace{-8mm}

Fig.~\ref{Fig:oscillograms} shows that the suppression of $\nu_e\to\nu_{\mu,\tau}$ oscillations is really significant in DNT with respect to CNT 
for the multi-GeV events with $E_\nu>5$\,GeV at the core-crossing trajectories. 

The main differences of the shown DNT result from the CNT one for the mantle-only crossing trajectories are:  
(1) larger values of $P_A$ in the lower part of shown ($\Theta_\nu$, $E_\nu$) area; 
(2) existence of the two bands (instead of one) with large values of $P_A$, which are separated by the band with small values of $P_A$, in the energy range $5~\text{GeV}<E_\nu<15$~GeV. 

For the core-crossing trajectories the differences between DNT and CNT results are even more dramatic: 
the picture derived in DNT has partly mirror dependence on the nadir angle with respect to the picture derived in CNT. 
Also the values of $P_A$ are essentially smaller in DNT with respect to CNT for $E_\nu>5$\,GeV.

The discussed differences should significantly effect fitting of the experimental data, using the simulated fluxes of atmospheric 
neutrinos~\cite{Honda:2011nf,Honda:2004yz}. This is a subject of future researches, which may include a detailed 
investigation of the oscillation channels $P_{\mu\mu}$ and $P_{\mu\tau}$.  
Implementation of new methods~\cite{1309.3176} would be also useful in further investigations.



In conclusion,  we have found the neutrino oscillogram driven by the 1-3 mixing in the matter of the Earth in the framework of the 
Democratic Neutrino Theory. This oscillorgam shows the significant suppression of $\nu_e\to\nu_{\mu,\tau}$ oscillations for 
the core-crossing trajectories of neutrinos with the energies $E_\nu>5$\,GeV.  
This is an important step on the way of accurate verification of this theory.

\section*{Acknowledgements} 

The author thanks Michele Maltoni and Gil Paz for useful comments. 
This work was supported in part by the US Department of Energy under the contract DE-SC0007983.

\end{document}